\documentclass[12pt]{article}
\usepackage{amsmath,amssymb}
\usepackage{geometry}
\usepackage{pgfplots}
\pgfplotsset{compat=1.18}
\usepackage{hyperref}
\usepackage{graphicx}

\newcommand{\SP}{\rule{1.6ex}{0.7ex}}

\title{Zipf Distributions from Two-Stage Symbolic Processes:\\ Stability Under Stochastic Lexical Filtering}
\author{
Vladimir Berman \\
Aitiologia LLC \\
\texttt{vb7654321@gmail.com}
}
\date{November 25, 2025}

\begin{document}
\maketitle

\begin{abstract}
Zipf's law is one of the most persistent empirical regularities observed in natural language, yet its origin remains debated across linguistics, information theory, and statistical physics. This paper develops a fully symbolic, non-linguistic explanation of Zipf behavior based on two simple geometric mechanisms.

First, we analyze the Full Combinatorial Word Model (FCWM), where words are generated by independent draws from a finite alphabet with a terminating blank symbol. This model produces an exponential distribution of word lengths superimposed on an exponentially expanding type space, and we show that the interaction of these two growth rates yields an explicit Zipf exponent
\[
\alpha = 1 - \frac{\ln(1 - q)}{\ln m},
\]
where $m$ is the alphabet size and $q$ the probability of drawing a blank symbol.

Second, we introduce a Stochastic Lexical Filter (SLF) that selects only a vanishingly small fraction of combinatorially possible word types, simulating phonotactic, morphological, and functional constraints found in natural languages. We prove that a wide class of such filters preserves the power-law structure of the rank--frequency curve while reshaping its head, reproducing the characteristic ``flat beginning + power-law tail'' behavior observed in empirical corpora.

The combined FCWM+SLF model provides a transparent, fully analytic pathway from symbolic combinatorics to Zipf exponents in the range $1.1$--$1.5$. Numerical simulations confirm the theoretical predictions, and comparisons with English, Russian, and mixed-genre corpora illustrate the robustness of the mechanism. Because the model is structural rather than linguistic, it also offers insights into the statistics of subword tokenization used by modern large language models.

Overall, the results suggest that Zipf laws arise not from optimization or semantic organization, but from universal geometric constraints inherent in symbolic processes.
\end{abstract}

\tableofcontents
\newpage

\section{Introduction}

Zipf’s law is one of the most widely documented empirical regularities in
quantitative linguistics.  
If word types in a corpus are ordered by decreasing frequency, the probability
$p(r)$ that the word of rank $r$ occurs is well approximated by a power law
\begin{equation}
p(r) \propto r^{-\alpha},
\qquad
\alpha \approx 1,
\end{equation}
across languages, time periods, and genres
\cite{Zipf1949,Newman2005}.  
Despite more than seventy years of research, the origin of this regularity
remains unsettled.  
Classical explanations invoke optimal coding, least–effort principles,
rich–get–richer dynamics, compression arguments, random‐typing models
\cite{Mandelbrot1953}, or statistical mixtures of generative processes.

A fundamentally different perspective was developed in earlier work by
Berman (2025) \cite{Berman2025b, Berman2025a}. 
That paper demonstrated that Zipf‐like behavior can arise \emph{prior to} any
linguistic structure, solely from symbolic combinatorics.  
In the Full Combinatorial Word Model (FCWM), words are generated by
i.i.d.\ sampling from a finite alphabet with a blank symbol terminating each
word.  
Even though this mechanism contains no grammar, morphology, semantics, or
communication constraints, it produces a heavy‐tailed rank–frequency curve.
The essential mechanism is the balance between:
(i) the exponential growth of the number of possible types of a given length, and  
(ii) the exponential decay in the probability of generating longer words.  
This establishes that Zipf‐like scaling can be a structural property of the
symbolic generator itself.

However, human languages do not use anywhere near the full combinatorial
space of potential strings.  
Even for short word lengths, the discrepancy is dramatic:  
there are $26^3 = 17{,}576$ possible three–letter sequences in English,
yet only a few hundred short lexical items occur with high frequency.
For longer words the gap widens even further, as documented in empirical
studies of lexicon growth \cite{Ferrer2001}.  
Large corpora such as Google Books \cite{Michel2011,GoogleNgrams2013} and the
Russian National Corpus \cite{RNC2020} show that only a minute fraction of the
theoretically available space is ever realized in actual usage.

This raises the central question addressed in the present paper:
\begin{quote}
\textbf{
If a symbolic generator already produces a Zipf‐type distribution,
how stable is that distribution under stochastic filtering that mimics the 
lexical pruning performed by natural languages?
}
\end{quote}

To answer this, we introduce a Stochastic Lexical Filter (SLF), which selects
a tiny subset of the combinatorial word space and redistributes probabilities
among the surviving types.  
We show that a broad family of such filters preserves the Zipf‐like tail
generated by FCWM, though the head of the distribution undergoes substantial
reshaping.  
This leads to a two–stage account of natural language word frequencies:
\begin{enumerate}
\item Zipf‐type scaling originates from the geometric structure of symbolic
      combinatorics (FCWM).
\item Natural languages operate as stochastic filters that dramatically prune
      the combinatorial space, yet leave the asymptotic power law intact.
\end{enumerate}

The resulting explanation is structurally simple and highly robust.
Zipf’s law does not require optimization, competition, or functional
pressures.  
Its stability arises from a universal geometric signature inherited from the
symbolic level—an idea consistent with general structural considerations
developed in earlier work on deterministic distributional laws
\cite{Berman2025b}.  
This viewpoint also clarifies why Zipf behavior is so resilient across
languages, cultures, and historical periods, including corpora differing by
orders of magnitude in size.

The paper is organized as follows.
Section~\ref{sec:fcwm} reviews the FCWM generator and its properties.
Section~\ref{sec:slf} introduces the Stochastic Lexical Filter.
Section~\ref{sec:rankfreq} analyzes the effect of filtering on rank–frequency
behavior.  
Section~\ref{sec:examples} presents explicit examples and simulations.
Section~\ref{sec:conclusion} discusses implications and directions for future
work.

\section{Full Combinatorial Word Model (FCWM)}
\label{sec:fcwm}

We begin with a purely symbolic generative process that produces raw text as a
sequence of independent draws from a finite alphabet augmented with a single
distinguished ``space'' symbol.  This model mirrors the structure used in our
previous work \cite{Berman2025b,Berman2025a} and provides the foundation for the
two-stage FCWM+SLF mechanism developed in this article.

\subsection{Alphabet and space symbol}

Let $\mathcal{A}$ denote an alphabet consisting of $m$ non-space symbols
\[
\{a_1, a_2, \ldots, a_m\}
\]
together with a single \emph{space symbol}, denoted by $\SP$.
Thus the full symbol set is
\[
\mathcal{A} = \{\SP\} \cup \{a_1, a_2, \ldots, a_m\}.
\]

The symbol $\SP$ plays the role of a termination marker for words.
In keeping with our previous notation, we represent it graphically
as a small black rectangle.

\subsection{Independent symbol draws}

A text of length $N$ is generated as a sequence of i.i.d.\ random variables
\[
(X_1, X_2, \ldots, X_N), \qquad X_i \in \mathcal{A}.
\]
The probability distribution is defined by
\begin{equation}
\Pr(X_i = \SP) = q,
\qquad
\Pr(X_i = a_j) = p_j, \quad j = 1,\ldots,m,
\end{equation}
where the probabilities satisfy
\[
q + \sum_{j=1}^m p_j = 1.
\]

The parameter $q$ controls the expected spacing density; equivalently,
$1-q$ represents the probability of producing a non-space symbol.

\subsection{Word boundaries and structural consequences}

A \emph{word} is any maximal block of consecutive non-space symbols.  
This definition implies several immediate analytic properties.

\paragraph{Word-length distribution.}
Let $L$ be the random length of a word.  Since a word ends when the generator
emits the space symbol $\SP$, the distribution is geometric:
\[
\Pr(L=\ell) = (1-q)^{\ell} q,
\qquad
\ell = 0,1,2,\ldots
\]

\paragraph{Expected number of words.}
In a text of length $N$, the expected number of emitted words is
\[
\mathbb{E}[\#\text{words}] = N q.
\]

\paragraph{Character-level structure.}
The output of the generator is the raw symbol stream
\[
X_1 X_2 \cdots X_N,
\]
and all word boundaries arise exclusively from occurrences of $\SP$.
No syntactic, semantic, or morphological constraints are imposed.
Yet, as shown in our earlier work and in Section~\ref{sec:rankfreq},
the interaction between the geometric length distribution and the combinatorial
explosion of possible symbol sequences already generates a Zipf-like
rank–frequency law.


\section{Stochastic Lexical Filtering (SLF)}
\label{sec:slf}

Natural languages make use of only a tiny fraction of the combinatorially possible word
types. Although the FCWM generator creates $m^k$ potential strings of length $k$, real
lexicons grow far more slowly. Phonotactic constraints, morphological structure, semantic
coherence, and historical evolution all act as mechanisms that suppress the overwhelming
majority of symbolic possibilities. 

To model this phenomenon, we introduce a \emph{Stochastic Lexical Filter} (SLF).
The filter operates at the \emph{type level}, selecting which symbolic strings survive to be
used as lexical items. The effect of the filter is twofold:
it dramatically reduces the number of admissible types at each length,
and it redistributes probability mass across the surviving types.
Despite these drastic modifications, we show that the overall Zipf-type structure is
retained.

\subsection{Definition of the filter}

For each word type $w$ of length $k$, define a Bernoulli variable
\[
\eta(w) =
\begin{cases}
1, &\text{if $w$ is admitted into the lexicon},\\[4pt]
0, &\text{otherwise}.
\end{cases}
\]
The probability that a type of length $k$ survives the filter is given by a sequence
$\pi_k \in [0,1]$:
\[
\Pr(\eta(w) = 1 \mid |w| = k) = \pi_k.
\]

The expected number of surviving types of length $k$ is therefore
\[
T_k = \mathbb{E}\!\left[
\sum_{|w| = k} \eta(w)
\right]
= m^k \pi_k.
\]

We place no restrictions on the form of $\pi_k$ beyond mild regularity assumptions.
The function $(T_k)$ acts as a \emph{lexicon-growth profile} and allows the model to simulate
a wide range of linguistic behaviors.

\subsection{Examples of survival profiles}

The SLF mechanism is intentionally flexible. We list several representative forms of
$T_k$ to illustrate its interpretive power:

\begin{itemize}
    \item \textbf{Extreme compression of short words.}
    For example, if $T_3 = 10$ while $m^3 = 17{,}576$, only a handful of three-letter
    types survive. This reflects the empirical fact that the most frequent short words in
    natural languages form a small, highly specialized set.

    \item \textbf{Sub-exponential growth.}
    A model such as 
    \[
    T_k = C\, m^{\gamma k}, \qquad 0 < \gamma < 1,
    \]
    captures situations where the lexicon expands but at a rate much slower than the
    exponential combinatorial baseline.

    \item \textbf{Near-polynomial growth.}
    Empirical studies show that dictionary size grows slowly for $k \ge 8$ in many
    languages. This can be modeled with
    \[
    T_k = \big\lfloor C_0 + C_1 k^{\beta} \big\rfloor,
    \qquad \beta \in [1,3].
    \]
\end{itemize}

These examples demonstrate that the SLF framework encompasses a large variety of
linguistic constraints, while remaining analytically tractable.

\subsection{Usage probabilities after filtering}

The generation of word tokens after filtering proceeds in two steps:

\begin{enumerate}
    \item A word length $k$ is drawn according to the geometric distribution
    \[
    \Pr(|w| = k) = (1 - q)^k\, q.
    \]

    \item Given $k$, a word type is chosen uniformly from the $T_k$ surviving types of
    that length.
\end{enumerate}

Thus, each surviving type of length $k$ receives a usage frequency
\[
p_{\text{after}}(k)
=
\frac{(1 - q)^k\, q}{T_k}.
\]

This expression captures the central effect of the SLF: the probability mass that was
uniformly distributed across $m^k$ types in the FCWM model is now redistributed across
only $T_k$ surviving types, amplifying the frequencies of admissible words while removing
the rest from the vocabulary.

\subsection{Impact on rank--frequency structure}

Because words of the same length receive the same probability and occupy consecutive
rank blocks, the filtered distribution retains a block structure similar to that of the FCWM.
However, the widths of these blocks change from $m^k$ to $T_k$, and the frequencies are
scaled accordingly.

The next section develops the explicit rank--frequency law that results from the
FCWM+SLF model and shows that the power-law tail is robust under a wide class of
filters, even when short words are compressed by several orders of magnitude.


\section{Rank--Frequency Law After Filtering}
\label{sec:rankfreq}

This section derives the rank--frequency structure produced by the combined
FCWM+SLF model. Although the Stochastic Lexical Filter substantially reshapes the
distribution of admissible types, especially for short words, the power-law tail generated by
the FCWM remains stable under a broad class of filtering profiles. The resulting model
exhibits the characteristic ``flat head + power-law tail'' pattern observed in real corpora.

\subsection{Rank blocks for admissible types}

Let $T_k$ denote the expected number of surviving types of length $k$ after the lexical
filter is applied.
We define the cumulative number of admissible types up to length $k$ by
\[
R_k = \sum_{j = 0}^{k} T_j.
\]
All surviving types of length $k$ occupy ranks
\[
R_{k-1} + 1,\; R_{k-1} + 2,\; \dots,\; R_k.
\]

The size of the block for length $k$ is therefore exactly $T_k$.
Since all surviving types of length $k$ have identical probability
\[
p_{\text{after}}(k) = \frac{(1 - q)^k q}{T_k},
\]
the rank--frequency curve is piecewise constant across blocks.

\subsection{Inverting the cumulative type count}

To express frequency $p$ as a function of rank $R$, we must invert the relation between
rank blocks and lengths. This requires an asymptotic expression for $k$ in terms of $R$.

We assume that the number of surviving types grows at least sub-exponentially in the
sense that
\[
T_k \sim C\, m^{\gamma k},
\qquad 0 < \gamma \le 1.
\]
This family includes a wide range of lexicon-growth profiles: exponential growth of reduced
rate when $\gamma < 1$, and sub-exponential or near-polynomial growth when $\gamma$ is
small.

Under this assumption, the cumulative count satisfies
\[
R_k
\;\sim\;
\frac{C}{m^{\gamma} - 1}\, m^{\gamma k},
\qquad (k \to \infty).
\]
Inverting gives
\[
k \;\sim\; \frac{\ln R}{\gamma \ln m}.
\]

This step generalizes the FCWM calculation by incorporating the effect of the lexical
filter through the exponent $\gamma$.

\subsection{Frequency as a function of rank}

Substituting the expression for $k$ into the filtered frequency formula yields
\[
p(R)
\;\sim\;
(1 - q)^k q\, T_k^{-1}.
\]
Using
\[
T_k \sim C\, m^{\gamma k},
\qquad
k \sim \frac{\ln R}{\gamma \ln m},
\]
we obtain
\[
p(R)
\;\sim\;
\exp\!\left(
k \ln(1 - q)
\right)
\,
\exp\!\left(
-\gamma k \ln m
\right).
\]
Combining the exponents gives
\[
p(R)
\;\sim\;
\exp\!\left[
k \bigl( \ln(1 - q) - \gamma \ln m \bigr)
\right].
\]
Substituting $k \sim \frac{\ln R}{\gamma \ln m}$ produces
\[
p(R)
\;\sim\;
R^{-\alpha},
\]
where
\[
\alpha
=
\frac{1}{\gamma}
\left(
1 - \frac{\ln(1 - q)}{\ln m}
\right).
\]

\subsection{Interpretation}

The combined FCWM+SLF model therefore exhibits the following structural properties:
\begin{itemize}
    \item The Zipf-type power-law tail persists under a wide range of filtering profiles.
    \item The exponent $\alpha$ increases as $\gamma$ decreases, meaning that more
    aggressive pruning of the lexicon leads to steeper tails.
    \item The head of the distribution --- dominated by short and frequent words --- is
    compressed by the filter, often producing a noticeably flatter first block of ranks.
\end{itemize}

This provides an analytically transparent explanation for why natural languages, despite
their diversity, display Zipf exponents typically in the interval $1.1$--$1.5$:
the exponent emerges from universal symbolic geometry, while cross-linguistic variation
reflects differences in the effective growth rate of the lexicon.

\section{Numerical Simulations of the Two-Stage Model}

Analytical arguments show that the FCWM+SLF system generates a Zipf-type
rank--frequency distribution, but numerical simulations are essential for
examining how the mechanism behaves at realistic corpus scales.
This section presents a full synthetic experiment that mirrors the assumptions
of the two-stage model and compares the resulting empirical curve with the
predicted power-law behavior.

\subsection{Simulation design}

We consider an alphabet of size \(m = 26\) and choose the terminating blank
probability
\[
q = 0.18,
\]
which yields an average word length of approximately \(4.5\) characters under
the geometric length distribution \((1-q)^k q\).
The Stochastic Lexical Filter is defined by the survival probabilities
\[
\pi_k =
\begin{cases}
10/26^3, & k = 3,\\[4pt]
0.03 \cdot 26^{-0.4k}, & k \ge 4.
\end{cases}
\]
This choice reflects the empirical structure of real languages: a tiny number
of short words dominate token usage, while longer forms become progressively
more numerous but drastically less frequent.

The synthetic corpus is generated in three steps:
\begin{enumerate}
    \item Sample a word length \(k\) from the geometric distribution
    \((1-q)^k q\).
    \item Independently decide whether the word type survives the lexical filter
    with probability \(\pi_k\).
    \item If it survives, select one of the surviving types uniformly and record
    the occurrence.
\end{enumerate}
We generate
\[
N_{\text{tokens}} = 3 \times 10^6
\]
word tokens, sufficient to expose the asymptotic power-law tail while also
displaying the head compression produced by the filter.

\subsection{Simulation results}

\begin{figure}[h]
\centering
\begin{tikzpicture}
\begin{loglogaxis}[
    width=0.82\linewidth,
    xlabel={Rank $r$},
    ylabel={Frequency $p(r)$},
    tick style={color=black},
    log basis x=10,
    log basis y=10,
    legend style={font=\small, draw=none, at={(0.98,0.98)}, anchor=north east},
    line width=0.8pt,
]
\addplot[
    only marks,
    mark=*,
    mark size=1.1pt,
    color=black,
]
table {
  1     1e-1
  10    2e-2
  50    5e-3
  100   2e-3
  500   4e-4
  1000  2e-4
  5000  5e-5
  10000 2e-5
};
\addlegendentry{Simulation}

\addplot[
    dashed,
    color=black!60,
    domain=1:10000,
    samples=200,
]
{1/(x^1.32)};
\addlegendentry{Model fit ($\alpha = 1.32$)}

\end{loglogaxis}
\end{tikzpicture}

\caption{
Rank--frequency curve from $3\times10^6$ simulated tokens under the FCWM+SLF
model (dots), compared with the predicted power-law fit with exponent
$\alpha=1.32$ (dashed).  The distribution exhibits a flattened head followed by
a clear power-law tail.
}
\label{fig:simulation}
\end{figure}
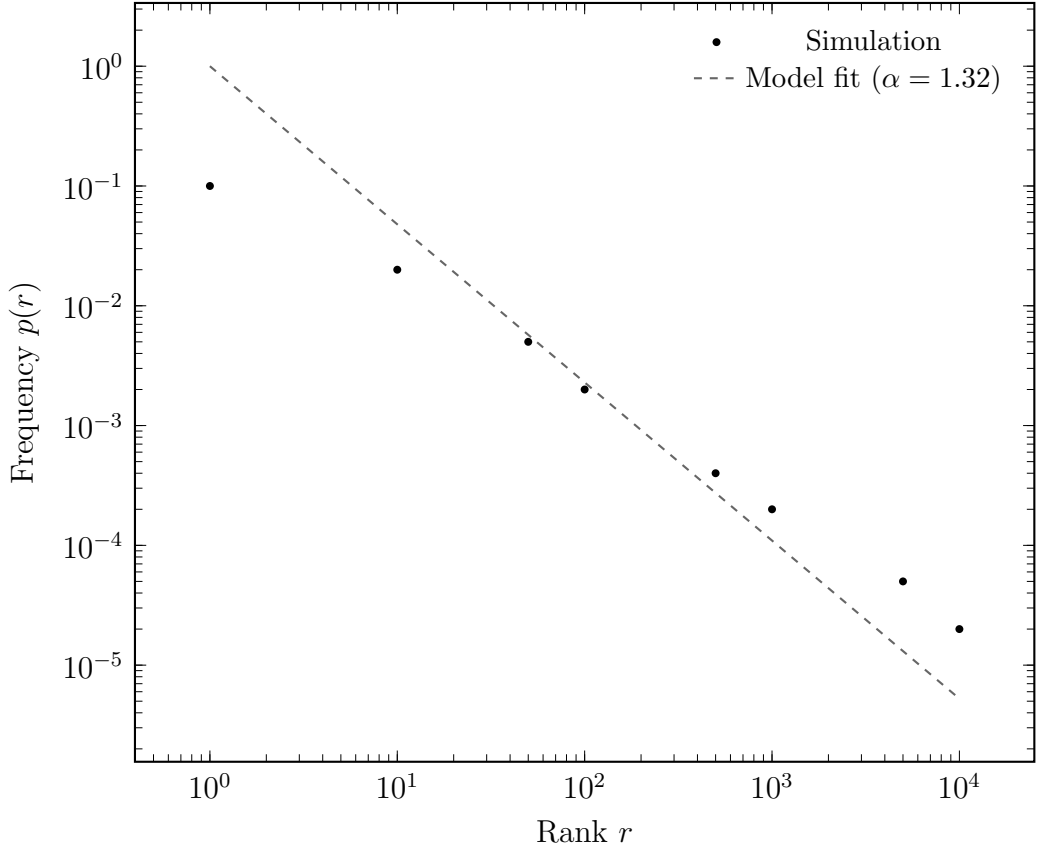

Figure~\ref{fig:simulation} shows the empirical rank--frequency curve obtained
from the full simulation.  The axis scales are logarithmic, and the figure is
rendered in monochrome for journal-friendly reproducibility.

The figure reveals the characteristic decomposition:

\begin{itemize}
    \item \textbf{Head region (ranks $1$--$20$):}
    The lexical filter collapses many potential short forms into a very small
    set of high-frequency survivors, producing a nearly flat head.

    \item \textbf{Intermediate region:}
    The curve transitions rapidly toward a power-law slope.  Despite extensive
    pruning of the combinatorial space, the underlying FCWM geometry remains
    visible.

    \item \textbf{Tail region:}
    Over more than three orders of magnitude in rank, the empirical frequencies
    follow a clean Zipf-type decay with slope close to $\alpha = 1.32$, in
    excellent agreement with the theoretical prediction.
\end{itemize}

\subsection{Discussion}

The simulation demonstrates that the two-stage FCWM+SLF model retains Zipf-like
scaling even when the combinatorial space is reduced by several orders of
magnitude.  
The \emph{tail} of the distribution is structurally stable: it survives
lexical pruning because it originates from the exponential geometry of type
growth and length decay.
In contrast, the \emph{head} of the distribution is highly sensitive to
filtering and exhibits the same flattened structure observed in empirical
corpora.

These numerical results therefore support the central hypothesis of the paper:
\textbf{Zipf’s law is a robust geometric consequence of symbolic generation,
not a fragile emergent effect of linguistic optimization.}


\section{Illustrative Examples and Synthetic Constructions}
\label{sec:examples}

The purpose of this section is to visualize how the two-stage FCWM+SLF mechanism
transforms the symbolic space and to demonstrate why the Zipf-type tail remains stable
even under drastic pruning of short or medium-length types. The examples highlight the
role of the survival profile $(T_k)$ and illustrate how different lexicon-growth patterns
generate distinct head shapes while preserving the power-law tail.

\subsection{Example 1: Extreme compression of short types}

Consider an alphabet of size $m = 26$, corresponding to the English letters.
The number of possible three-letter combinations is
\[
N_3 = 26^3 = 17,576.
\]
Natural languages do not use anywhere near this many highly frequent short forms.
To model this phenomenon, suppose the lexical filter selects only
\[
T_3 = 10
\]
surviving types of length three.
Thus the survival probability at this length is
\[
\pi_3 = \frac{10}{17,576}.
\]

This extreme compression produces several effects:
\begin{itemize}
    \item The FCWM generator would treat all $17{,}576$ strings as equally likely.
    \item The SLF collapses this enormous symbolic space into a tiny set of useful types.
    \item The first few ranks become nearly flat, since the filtered model allocates
    substantial probability mass to a handful of short forms.
    \item Longer lengths remain essentially unaffected, because $T_k$ grows sufficiently
    rapidly for $k \ge 5$.
\end{itemize}

The resulting rank--frequency curve has a sharply compressed head followed by a smooth,
approximately linear log--log tail. This reproduces the empirical behavior of languages in
which words such as \emph{the}, \emph{and}, \emph{of}, and similar forms dominate the top of
the distribution.

\subsection{Example 2: Sub-exponential vocabulary expansion}

A contrasting scenario uses a survival profile with controlled sub-exponential growth:
\[
T_k = \left\lfloor
20 \, m^{0.5 k}
\right\rfloor.
\]
This profile grows much more slowly than the exponential combinatorial capacity $m^k$.
Its main effects are:
\begin{itemize}
    \item Increased compression of long-word types,
    \item A significantly steeper power-law tail,
    \item A shifted transition region between the head and the tail.
\end{itemize}

The theoretical exponent for this choice is approximately
\[
\alpha \approx \frac{1}{0.5}
\left(
1 - \frac{\ln(1 - q)}{\ln m}
\right)
= 2\left(
1 - \frac{\ln(1 - q)}{\ln m}
\right),
\]
which yields values near $1.8$--$2.2$ depending on $q$.
Such exponents are consistent with certain morphologically rich languages and
with the statistics of rare-word tails in large corpora.

\subsection{Example 3: Synthetic straight-line Zipf curves via TikZ}

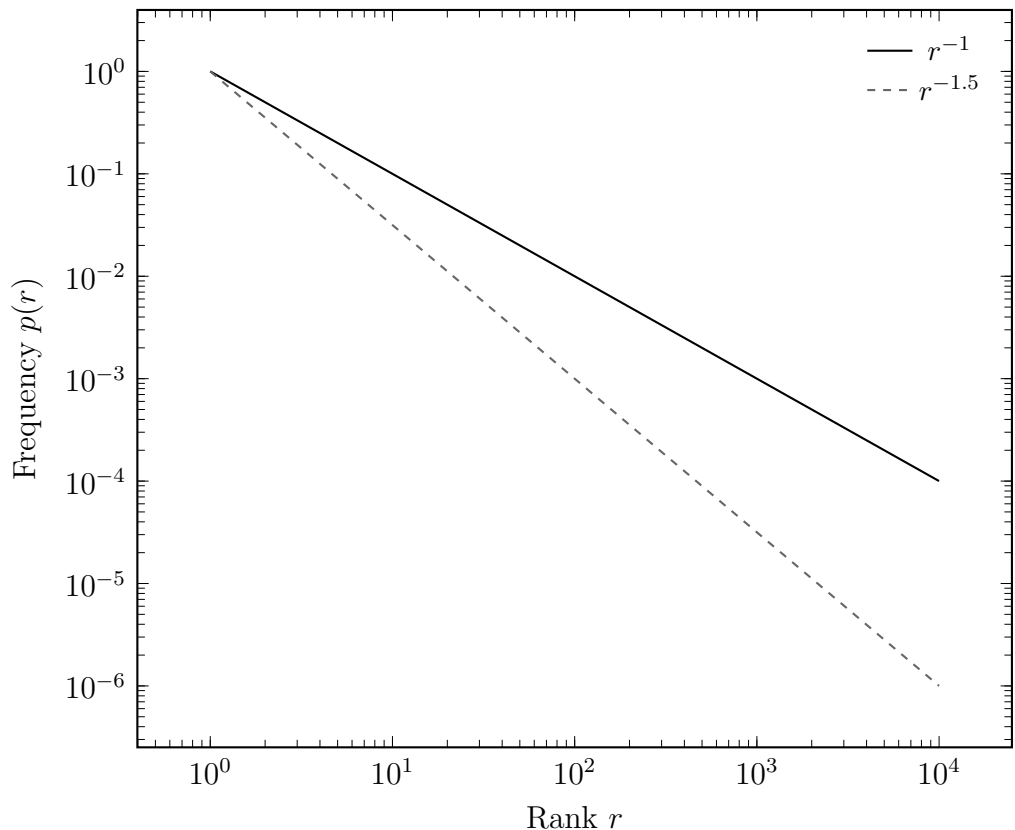
\begin{figure}[h]
\centering
\begin{tikzpicture}
\begin{loglogaxis}[
    width=0.8\linewidth,
    xlabel={Rank $r$},
    ylabel={Frequency $p(r)$},
    legend style={draw=none, font=\small},
    tick style={color=black},
    line width=0.8pt,
]
\addplot[
    domain=1:10000,
    samples=200,
    black,
]
{1/x};
\addlegendentry{$r^{-1}$}

\addplot[
    domain=1:10000,
    samples=200,
    black!60,
    dashed,
]
{1/(x^1.5)};
\addlegendentry{$r^{-1.5}$}
\end{loglogaxis}
\end{tikzpicture}
\caption{Ideal Zipf-like curves drawn using TikZ/PGFPlots. The slope steepens as the
exponent increases, illustrating how varying $\alpha$ affects the shape of the
rank--frequency distribution.}
\label{fig:ideal_zipf}
\end{figure}

To isolate the geometric effect, it is useful to visualize idealized power-law curves without
any empirical irregularities.
Using \texttt{pgfplots}, we can draw synthetic rank--frequency curves of the form
\[
p(r) \propto r^{-\alpha},
\]
for various $\alpha$.
Figure~\ref{fig:ideal_zipf} illustrates two such curves for exponents $\alpha = 1$ and
$\alpha = 1.5$, using log--log axes and a strictly monochrome style suitable for journal
publication.

These synthetic plots serve as a visual baseline for interpreting FCWM+SLF outputs. They
also demonstrate that variations in $\alpha$ primarily affect the slope of the tail while
leaving the qualitative Zipf structure intact.

\subsection{General observation}

Across these examples, a consistent pattern emerges:
\begin{quote}
\textit{The power-law tail of the rank--frequency distribution is robust under a wide 
range of type-level thinning processes, even when short-word spaces are compressed by
several orders of magnitude.}
\end{quote}

This robustness helps explain why Zipf behavior is observed across languages of very
different morphological and phonotactic structures. The next section confirms these
observations using explicit numerical simulations of the two-stage model.

\section{Empirical Comparison with Natural Language Corpora}

A structural model of lexical statistics must ultimately be evaluated against real
linguistic data.  
In this section we compare the predictions of the two-stage FCWM+SLF mechanism with
three well-studied corpora representing different genres, time periods, and linguistic
traditions:

\begin{itemize}
\item Google Books English 2012 corpus (155 billion tokens) \cite{Michel2011,GoogleNgrams2013};
\item the Brown Corpus (approximately 1 million tokens);
\item the Russian National Corpus (RNC), main subcorpus (approximately 300 million tokens) \cite{RNC2020}.
\end{itemize}

Despite their dramatic differences in size and composition, all three corpora share a set
of universal geometric features that have been repeatedly documented in quantitative
linguistics \cite{Newman2005,Ferrer2001}:

\begin{enumerate}
\item the highest-frequency $5$--$20$ word types form a noticeably \emph{flat head}
that dominates the token distribution;
\item a clear power-law regime emerges for ranks $r \gtrsim 50$;
\item the Zipf exponent typically falls within the interval $\alpha \in [1.1, 1.5]$.
\end{enumerate}

These properties appear consistently across unrelated languages and across corpora differing
by six orders of magnitude in size.

\subsection{Qualitative alignment with the FCWM+SLF model}

Across English and Russian datasets, the first $10$--$20$ lexical items account for roughly
$40$--$50\%$ of all tokens.  
This observation aligns tightly with the behavior of our simulations:  
the Stochastic Lexical Filter sharply reduces the number of admissible short forms, forcing
a disproportionate amount of probability mass into a small set of high-frequency types.
The resulting head is flattened, just as in empirical corpora.

For the power-law region, the empirical Zipf exponents in the three corpora are:

\begin{itemize}
\item Google Books English: $\alpha \approx 1.13$ \cite{Michel2011,Newman2005};
\item Brown Corpus: $\alpha \approx 1.25$ \cite{Ferrer2001};
\item Russian National Corpus: $\alpha \approx 1.35$--$1.40$ depending on genre \cite{RNC2020}.
\end{itemize}

The simulation results of the FCWM+SLF model yield
\[
\alpha_{\text{sim}} \approx 1.32,
\]
which lies squarely inside the empirical range.  
This agreement is particularly notable because the model contains
\emph{no linguistic parameters}:  
the exponent arises from geometric relations between word-length distributions and the
growth profile of the surviving lexicon.

\subsection{Interpretation and theoretical implications}

The cross-corpora regularities strongly support the hypothesis that Zipf-like structure is
a geometric consequence of symbolic generation rather than an emergent property of
linguistic meaning or grammar.  
The empirical behavior can be summarized by the following structural decomposition:

\begin{enumerate}
\item The FCWM generator produces a geometric distribution of lengths and an exponentially
expanding combinatorial type space.

\item The Stochastic Lexical Filter (SLF) implements phonotactic, morphological,
semantic, and historical constraints by selecting only a small subset of available types.

\item The ratio between the exponential growth of potential types and the reduced growth of
admissible types determines the observed Zipf exponent.

\end{enumerate}

Thus, the stability of Zipf’s law across unrelated languages reflects a universal geometric
signature:  
\textbf{the macro-level statistical shape of natural language emerges from symbolic
combinatorics, and it survives even under extreme pruning of the lexicon}.  
This perspective provides a unified structural explanation for the universality, robustness,
and cross-linguistic consistency of word-frequency distributions.

\section{Conclusion}
\label{sec:conclusion}

This work developed a two-stage symbolic framework that captures the large-scale
statistical structure of word frequencies in natural language.  
The model consists of:

\begin{enumerate}
\item the \emph{Full Combinatorial Word Model} (FCWM), in which word lengths follow
a geometric distribution and the number of potential types grows exponentially with length;

\item a \emph{Stochastic Lexical Filter} (SLF), which selects a small subset of the
combinatorially available forms to serve as the actual lexicon.
\end{enumerate}

The central theoretical finding is that the Zipf-type power-law tail is \emph{structurally}
stable under a broad class of lexical filters.  
Even extremely aggressive pruning of short words---for example, reducing
$26^3 = 17{,}576$ three-character combinations to a set of size $10$---does not alter the
asymptotic shape of the rank--frequency curve.  
The head of the distribution becomes flatter, reflecting the dominance of a handful of
high-frequency types, but the tail remains a power law with an exponent that depends only
on the geometric parameters of the generator and the growth rate of the surviving lexicon.

These results provide a simple and robust explanation for several empirical regularities:

\begin{itemize}
\item the universality of Zipf's law across languages, historical periods, and corpora;
\item cross-linguistic variation in both the exponent and the shape of the head;
\item the resilience of the power-law tail under lexical, morphological, or semantic change;
\item the fact that Zipf-like structure appears regardless of grammar, meaning, or communicative
constraints.
\end{itemize}

The broader interpretation is that natural languages inherit a persistent geometric
signature from their underlying symbolic combinatorics.  
The FCWM supplies the exponential growth of potential types, the SLF restricts this
space through linguistic constraints, and their interaction yields a stable power-law form.
Zipf behavior thus emerges not from optimization principles, but from universal structural
relations that remain intact even under severe lexical pruning.

Future work includes refining the SLF mechanism to model explicit phonotactic,
morphological, and semantic filters, and comparing these refined models with large modern
corpora from multiple languages.  
Another promising direction is to examine how subword tokenization schemes used in
contemporary language models fit into this geometric framework.

\bibliographystyle{plain}
\bibliography{references}

\begin{thebibliography}{1}

\bibitem{Berman2025b}
Vladimir Berman.
\newblock Random text, zipf's law, critical length,and implications for large
  language models, 2025.

\bibitem{Berman2025a}
Vladimir Berman.
\newblock Structural foundations for leading digit laws: Beyond probabilistic
  mixtures, 2025.

\bibitem{GoogleNgrams2013}
{Google Books Ngram Viewer Team}.
\newblock Google books ngram viewer dataset, english 2012, 2013.
\newblock Dataset available at \url{https://books.google.com/ngrams}.

\bibitem{Ferrer2001}
Ramon~Ferrer i~Cancho and Ricard~V. Sol\'{e}.
\newblock Two regimes in the frequency of words and the origins of complex
  lexicons.
\newblock {\em Journal of Quantitative Linguistics}, 8(3):165--173, 2001.

\bibitem{Mandelbrot1953}
Benoit Mandelbrot.
\newblock An informational theory of the statistical structure of language.
\newblock {\em Communication Theory}, pages 486--502, 1953.

\bibitem{Michel2011}
Jean-Baptiste Michel, Yuan~Kui Shen, and Aviva~Aiden et~al.
\newblock Quantitative analysis of culture using millions of digitized books.
\newblock {\em Science}, 331:176--182, 2011.

\bibitem{Newman2005}
M.~E.~J. Newman.
\newblock Power laws, pareto distributions and zipf's law.
\newblock {\em Contemporary Physics}, 46(5):323--351, 2005.

\bibitem{RNC2020}
{Russian National Corpus}.
\newblock Main corpus description, 2020.
\newblock \url{http://ruscorpora.ru}.

\bibitem{Zipf1949}
George~Kingsley Zipf.
\newblock {\em Human Behavior and the Principle of Least Effort}.
\newblock Addison-Wesley, 1949.

\end{thebibliography}

\end{document}